\documentclass[a4paper]{article}

\usepackage{icrc2013}

\title{Commissioning and initial performance of the H.E.S.S. II drive system}

\shorttitle{The H.E.S.S. II drive system}

\authors{
P. Hofverberg$^{1}$,
R. Kankanyan$^{1}$,
M. Panter$^{1}$,
G. Hermann$^{1}$,
W. Hofmann$^{1}$,
C. Deil$^{1}$,
F. Ait Benkhali$^{1}$,
for the H.E.S.S. Collaboration.
}

\afiliations{
$^1$ MPI-K Heidelberg, Germany \\
}

\email{petter.hofverberg@mpi-hd.mpg.de}

\abstract
{
  The H.E.S.S. observatory was recently extended with a fifth telescope located at the center of the array - H.E.S.S. II. With a reflector roughly six times the area of the smaller telescopes and four times more pixels per sky area, this new telescope can resolve images of particle showers in the atmosphere in unprecedented detail and explore the $\gamma$-ray sky in the poorly studied regime around a few tens of Giga electron-volt.

  H.E.S.S. II has been equipped with a high-performance drive system that can deliver the high torque necessary to accelerate and slew the 600 tonnes telescope while keeping a good tracking accuracy. A modular design with a high degree of redundancy has been employed to achieve stability of operation and to ensure that the telescope can be moved to a safe position within a short period of time. Each axis is driven by four 28~kW servo motors which are pair-wise torque-biased and synchronized through a state of the art Programmable Logic Controller (PLC). With this system, a fast repositioning and a minimal settling time has been achieved - crucial when studying transient sources such as $\gamma$-ray bursts which are a prime target for this telescope.

  This contribution will report on the successful commissioning of the H.E.S.S. II drive system in the first half of 2012 at the H.E.S.S. site in Namibia. The technical implementation and the performance of the drive system will be presented.
}

\keywords{VHE $\gamma$-ray astronomy, telescope drive systems}

\begin{document}
\maketitle

\section{Introduction}
Ground based $\gamma$-ray astronomy is now an established branch of high energy astronomy with a large variety of source classes and close to 100 detected sources. The success was built upon the detection technique with \emph{imaging atmospheric Cherenkov telescopes} (IACTs) \cite{hegra}\cite{hess}, where Cherenkov light emitted from the atmospheric shower of particles created by the primary $\gamma$-ray is collected by a reflector and focused onto a pixelised camera. Using this approach, the direction of the primary $\gamma$-ray can be reconstructed geometrically from the image of the shower in multiple telescopes. Typically, this method gives a point spread function (PSF) of $\mathcal{O}(0.1^{\circ})$. The requirements on the tracking precision for IACTs is therefore more relaxed than for other optical telescopes which usually require sub arcsec resolution.

IACTs can only operate at full performance during clear moonless nights which severely limits the amount of observation time which is available each year. Each source can also only be effectively observed around its culmination due to the thicker atmosphere and consequently larger absorption of Cherenkov light at large angles from zenith. IACTs therefore need fast drive systems to minimize the loss of, the already limited, observation time caused by the frequent repositionings between sources.

The newly constructed H.E.S.S. II telescope is extending the old H.E.S.S. array with a fifth larger, 600~m$^2$ reflector area, telescope at the center of the array. One of the primary targets for H.E.S.S. II is transient sources like $\gamma$-ray bursts (GRBs), where a fast follow-up after the trigger is crucial to maximize the chance of detection \cite{fermigrb}.

\section{The H.E.S.S. II telescope}
\begin{figure*}[!t]
\centering
\includegraphics[width=\textwidth]{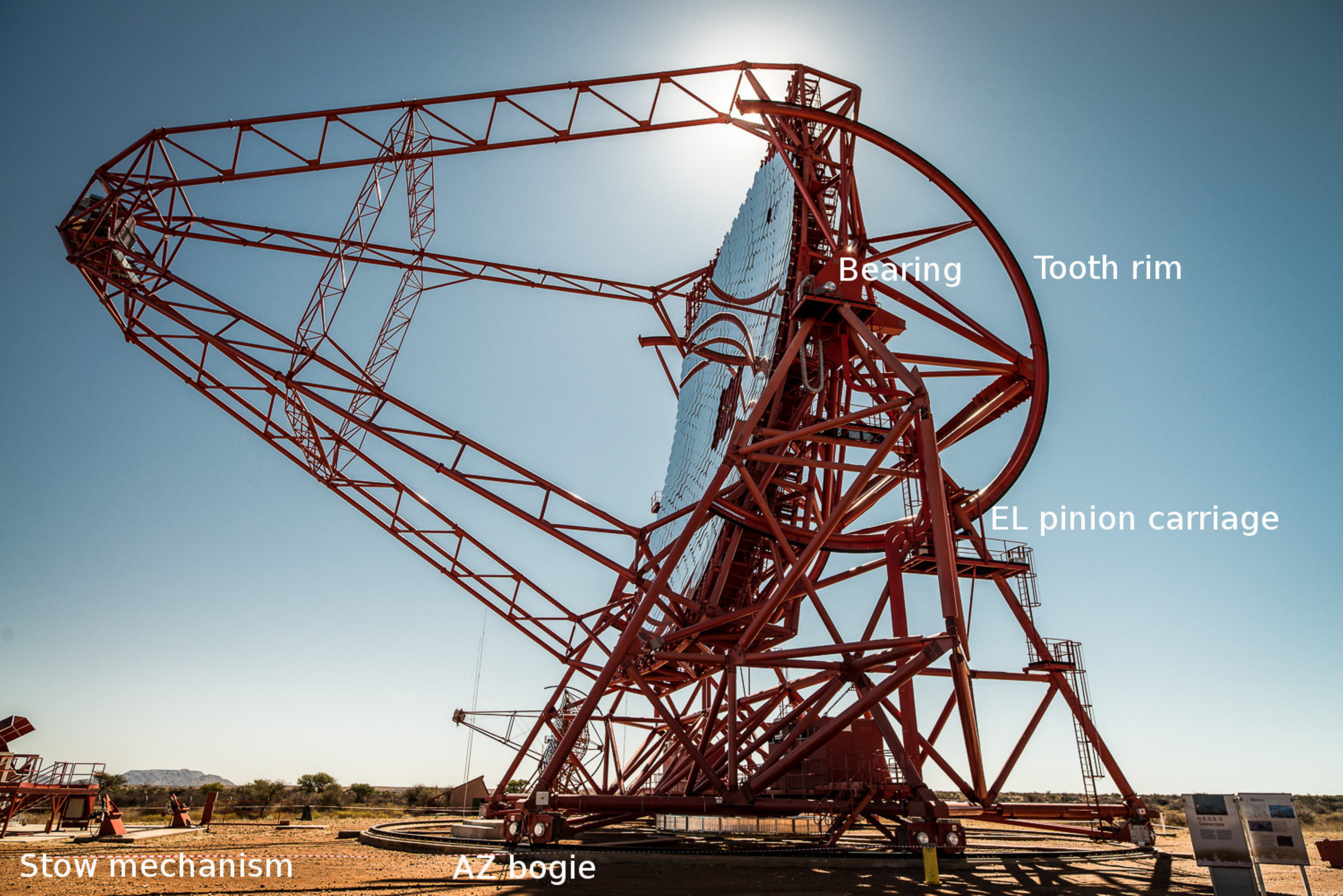} 
\caption{The H.E.S.S. II telescope with relevant parts marked out. The telescope is pointing in the south direction. One of the smaller telescopes is visible in the background. Photo by C.\,F{\"o}hr.}
\label{fig:telescope}
\end{figure*}

The H.E.S.S. II telescope is equipped with an altitude-azimuth mount and can thus be rotated around two orthogonal axis, i.e the vertical azimuth axis (AZ) and the horizontal elevation axis (EL). The complete telescope can be seen in Fig.~\ref{fig:telescope}.

The rotation around the AZ axis is realized through a wheel/rail system in connection with the central guide bearing (pintle bearing). The (flat) rail has a radius of 18~m. The mount is supported by six bogies, each with two conically shaped wheels ($\O$\,80~cm), which takes up to full weight of the telescope ($\sim$600~tonnes). 
The rotation around the EL axis is realized through two bearing units, each supported by a A-shaped framework of the mount. The bearing units are flange-connected laterally on to a dish support structure which serves as the interface between the dish and the mount. Two circular arcs on the dish support structure (of radius 12~m) are equipped with tooth-racks which are engaged by the EL drive mechanism.

The dish is a stiff spatial framework with a height of 32~m and width of 24~m. The side facing the camera has an approximately parabolic shape and is made up of 5x5 rectangular surfaces which serves as the fundaments for the mirror support segments, each holding mirror tiles of hexagonal shape with a width (flat to flat) of 0.9~m.
A quadropod supports the 3~ton camera in the focus of the reflector, at a distance of roughly 36~m. It is mounted to the corners of the dish through four legs designed to minimize the shadowing of the reflector. 

Two mechanical hooks are used to secure the telescope at the park position in south when the telescope is not used. In this condition, the camera can be unloaded and placed in a shelter by a fully automated (un)loading system.

\section{The drive system}

\subsection{Overview}
\begin{figure}[!t]
\centering
\includegraphics[width=0.4\textwidth]{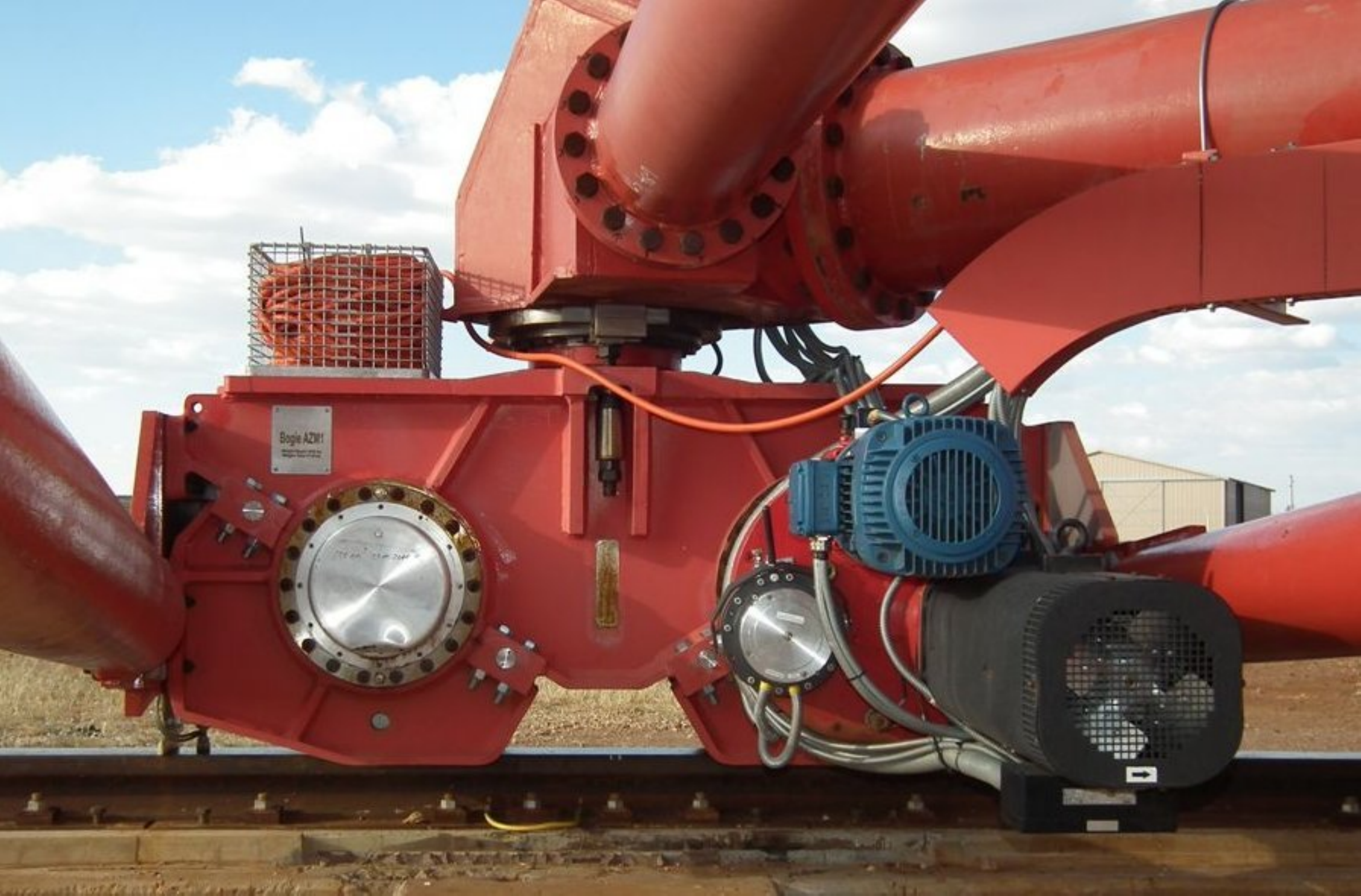}
\includegraphics[width=0.4\textwidth]{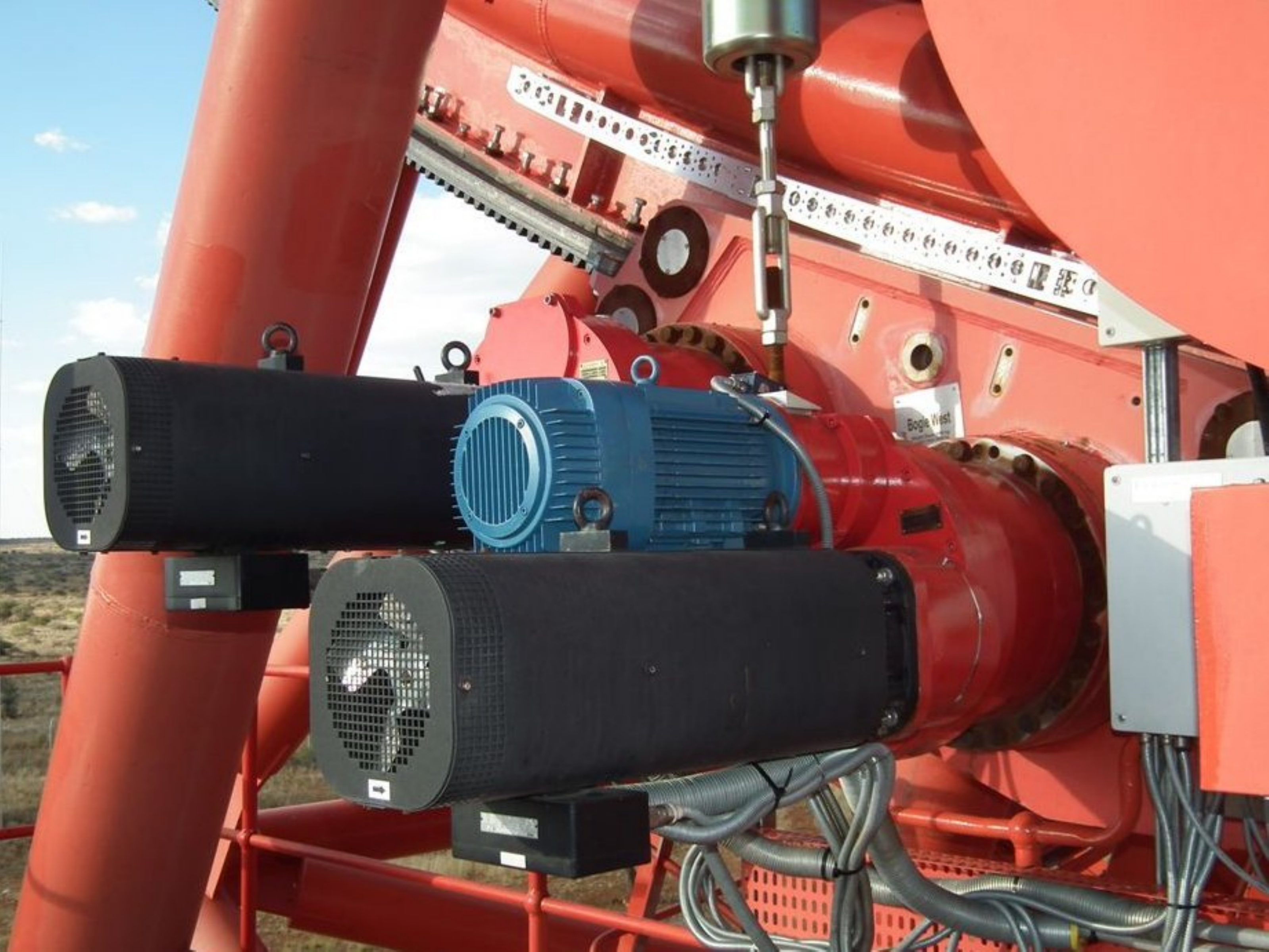}
\caption{\emph{Top} An AZ bogie. The servo motor (black) and emergency motor (blue) are engaging with the gear box through two drive shafts. The holding brake is visible just to the left of the servo motor. The bogie is rolling on the rail, visible in the lower part of the figure, via two conically shaped wheels (not visible). \emph{Bottom} The pinion carriage on the west side of the telescope. Two pinion wheels inside the carriage engages with the tooth-rack. Each pinion wheel is driven by a servo motor (black) via a gear set. One gear set is additionally equipped with an emergency motor (blue). Photos by F. van Greunen.}
\label{fig:driveunits}
\end{figure}

In AZ, four out of six bogies are equipped with a servo motor which engage with one of the bogie wheels through a planetary gear set. In EL, the drive is effected via the two toothed rims. Each tooth rim engage with a pair of drive pinions ($\O$\,36.6~cm) which in turn are driven by planetary gear sets equipped with servo motors. The pinions and the gear boxes are installed in two pinion carriages that are guided along the toothed rims by means of guide rollers and are supported on the AZ structure by means of flexure columns. A gear set equipped with a servo motor is referred to as a \emph{drive unit} and there are thus 4 drive units on each axis. One AZ and two EL drive units are shown in Fig.~\ref{fig:driveunits}.

The drive control is realized through a programmable logic controller (PLC) which provides set-point velocities to servo inverter modules. The control loop is closed by incremental encoders which gives information about the position and velocity of the telescope and motors. To add a second level of safety, position switches are used to restrict the movement of the telescope axes and are read in through safety input terminals to a safety PLC.

Tab.~\ref{tab:parameters} summarizes the most important characteristics of the telescope and drive system.\\

\begin{table}[h]
\begin{center}
\begin{tabular}{|l|c|}
\hline 
Parameter & Value\\ \hline
Acceleration (EL,AZ) & 0.5\,$^{\circ}/\mathrm{s}^2$ \\
Maximum velocity (EL,AZ) & 1.67\,$^{\circ}/\mathrm{s}$ \\
Maximum motor torque (EL,AZ) & $4\times170$\,Nm \\
EL Brake torque & $4\times140$\,Nm \\
AZ Brake torque & $4\times170$\,Nm \\
EL transmission ratio & 7875 \\
AZ transmission ratio & 5400 \\
EL moment of inertia & $7.67\mathrm{E}7~\mathrm{km}\cdot \mathrm{m}^2$ \\
AZ moment of inertia & $9.81\mathrm{E}7~\mathrm{km}\cdot \mathrm{m}^2$ \\
EL motion range & $-32^{\circ}$ to $175^{\circ}$ \\
AZ motion range & $78.5^{\circ}$ to $638.5^{\circ}$ \\
\hline
\end{tabular}
\caption{Parameters of the H.E.S.S. II telescope and drive system.}
\label{tab:parameters}
\end{center}
\end{table}

\subsection{Drive units}
Each gear set (Schafer, type GSP3-445-120-P-H-F265, gear ratio 119.43) has two separate drive shafts. One shaft on each gear set is equipped with a servo motor (AMK, type DH13-150-4-IOF), having a nominal power of 28~kW and a standstill torque of 170~Nm. The position and velocity of the motor shaft is measured by means of an incremental motor encoder (Heidenhain, ERN1380) which provides a 10~bit accuracy. An external holding brake (Stromag, NFF 16) is attached to the same drive shaft providing a brake torque of 170~Nm which allows the telescope to be secured at any position up to wind speeds of roughly 100~km/h. The brakes are normally controlled by the PLC but can also be forced open by an operator by means of a handle in case of a brake or PLC failure. 
The position of the brake disc is monitored by means of an integrated micro-switch which allows to compare the commanded and actual state of the brake. 

The second drive shaft on two of the gear sets on each axis are additionally equipped with a \emph{emergency motor} (Demag, KBA 125B 12/2). The emergency motors are normally disengaged from the gear drive shaft but can be engaged by means of a handle on top of the gear box when needed. The state of the emergency motor, i.e engaged or disengaged, is monitored by switches which prevents an operator to use the servo system while an emergency motor is engaged. The emergency motor operates with an internal brake.\\
\vspace{1.5mm}

\subsection{Drive control}
The direction and velocity with which a motor is running is set by supplying the motor with a three phase current with a specific phase and amplitude. This is provided by the servo drives. A servo drive is composed of a power supply module (AMK, KE60) and an inverter module (AMK, KW60). The supply and inverter modules each output 60~kW but can be overloaded to 120~kW for up to a minute. There are in total 8 inverter modules and 4 supply modules (each supplying power for an inverter module pair) distributed evenly in two drive cabinets for redundancy. A micro-computer inside the inverter module implements a position and velocity control loop, and calculates the instantaneous value of the required motor torque from the set-point motor velocity and the actual motor position and velocity. 

Motors are pair-wise torque biased against each other to reduce backlash in the gear train. The motor pre-tension, i.e the difference in torque between the driving and the braking motor in each torque biased motor pair, is 30~Nm. With this pre-tension, no measurable backlash could be detected and must thus be $<$1~arcsec.

The energy stored in the system (as kinetic energy of the moving axes) is transferred back to the supply modules when an axis is decelerating, so called regenerative braking. If more energy is transferred back than what can be stored, the remaining energy is released as heat through braking resistors. Uninterruptable power supplies (UPS) are used to backup all 24~VDC control logic and power supply. In this way, a controlled stop of the telescope can be performed even in the event of a power loss.

An embedded PC (Beckhoff, CX1020-0121) running Windows XP and TwinCAT IEC61131 PLC in parallel is used as a control computer. No rotating parts are used by the PC to increase the mean time before failure. A setup of two identical PCs is used, where a switch between the two can be done within minutes in case of a failure of the primary unit.

The emergency drive system can be operated from a switch box located on the telescope from where an operator can choose the velocity and direction of motion of each axis. LEDs on the box indicates the current direction of motion and the correct direction of motion for going towards the park position.

The telescope axes positions are provided by three position encoders (Heidenhain, ROC 417) located inside the bearings of the telescope axes: two in EL and one in AZ. The two EL encoders are continuously monitored for consistency. The encoders are absolute, single-turn encoders with a resolution of 17 bits which corresponds to a precision of roughly 10~arcsec. 

\section{Performance of the drive system}

\subsection{Tracking accuracy}
The H.E.S.S. II telescope is equipped with an Alt-Az mount and must therefore synchronize the movement of the EL and AZ axes to track a source on the celestial sphere. The pointing accuracy of the telescope, i.e the accuracy with which the optical axis of the telescope can be aligned with the direction to the source, depends on: i) the accuracy with which the drive system can track a target according to the shaft encoders (the \emph{tracking accuracy}), ii) the calibration accuracy of the shaft encoders, and iii) the mechanical bending and mis-alignment of the telescope structure. The two latter effects are compensated for offline using an empirical pointing model and is the subject of a future paper. The tracking accuracy depends solely on the performance of the drive system and should prefferably be significantly better than the PSF of the telescope to minimize blurring of sources. Since the projected velocity of the source on the telescope coordinate system depends strongly on the EL angle, the drive system must be able to ensure this tracking accuracy over a large variety of axis velocities.

\begin{figure}[t]
\centering
\includegraphics[width=0.4\textwidth]{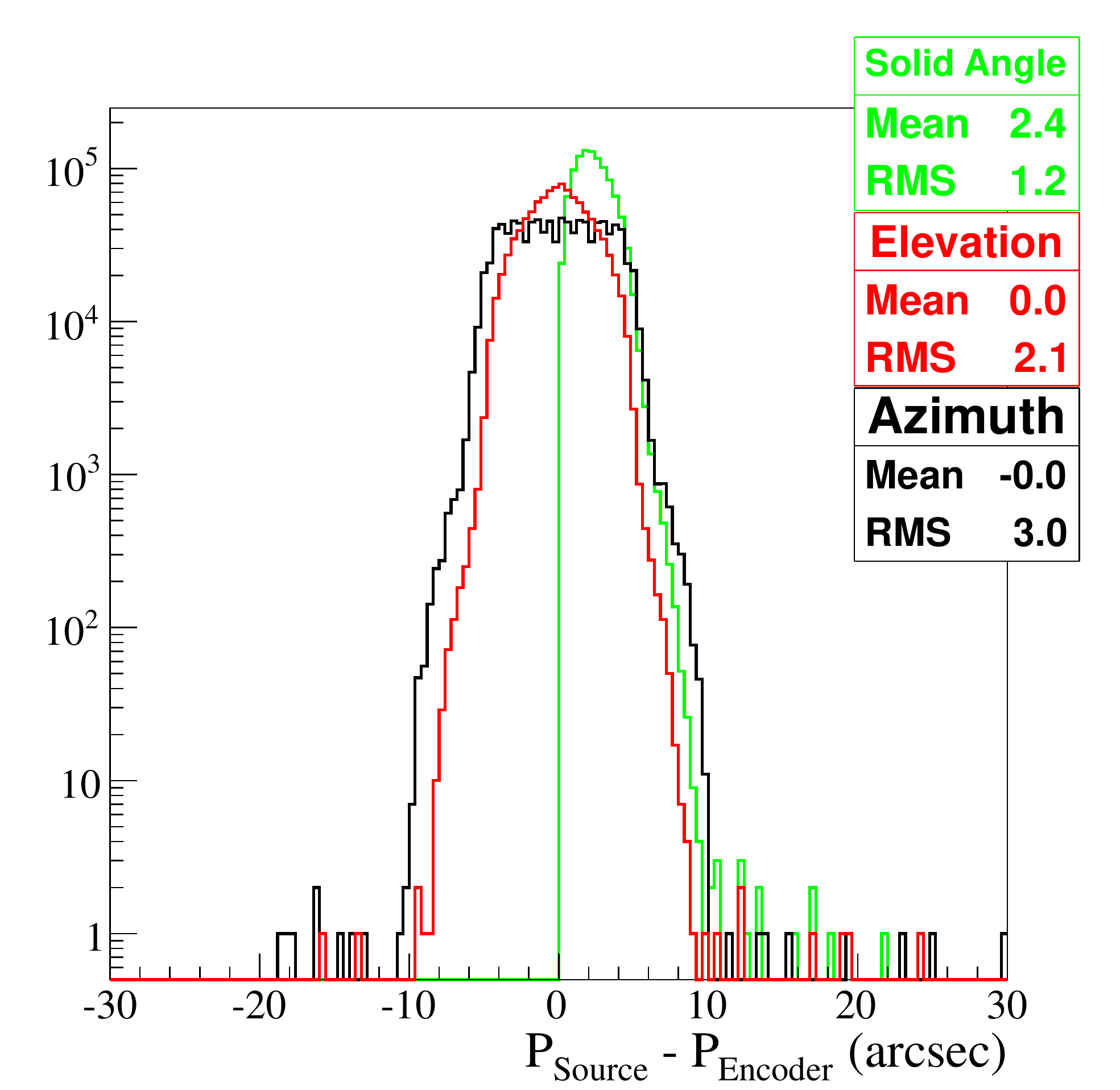}
\caption{The tracking accuracy of the CT5 telescope for $\sim$300~h of observations.}
\label{fig:trackingaccuracy}
\end{figure}

The tracking accuracy of the H.E.S.S. II telescope has been investigated using data from observations made during the first months of operation. Fig.~\ref{fig:trackingaccuracy} shows the distribution of the tracking accuracy for each axis and for the solid angle for roughly 300~h of tracking.  As shown, the drive system provides a very stable tracking accuracy, with a mean (solid angle) accuracy of $\sim$2.4~arcsec. This is several order of magnitudes smaller than the PSF of the H.E.S.S. II telescope, and the tracking accuracy can therefore safely be neglected in the offline analysis of H.E.S.S. II data.

\vspace{1.5mm}
\subsection{Time on target}
The time to reposition to a new position on the sky, the \emph{time on target},
depends on: i) the angular distance between the telescope pointing position and the new target, ii) the acceleration and velocity of the telescope axes, and iii) the settling time. The angular distance the telescope has to move to a target can be minimized in the design of the telescope, as has been done for H.E.S.S. II. Firstly, the telescope has been prepared for doing observations in \emph{reverse}, i.e allowing the EL axis to go beyond the zenith position ($90^{\circ}$). Secondly, the motion range in AZ has been made considerable larger than one full lap, with an overlap from east to west over the south position. In this manner, any position on the sky can be reached in 3 or 4 different ways (with reverse observations enabled), thus minimizing the repositioning distance to a target. Indeed, Monte-Carlo simulations show that such an overlap in AZ reduces the time on target by 25\% (8\%) with (without) reverse observations enabled, compared to having a minimal overlap in AZ of $25^{\circ}$.

The maximum velocity and acceleration of the telescope was fixed to the numbers given in Tab.~\ref{tab:parameters} during the commissioning of the telescope, and was constrained by the maximum motor torque, the performance of the feedback control loops and the flexibility of the steel construction. An increase of the acceleration and velocity of the AZ axis is being considered. The settling time is defined as the time required to go from full speed to a stable tracking of the source. By optimizing the feedback control loops in the PLC, a very low settling time of $\sim$5~s has been achieved. The settling time can however be eliminated completely by starting observations as soon as the source is in the field of view (FoV) of the camera. This technique will be used, together with reverse observations, for cases where a minimal repositioning time is required such as GRBs.

\begin{figure}[t]
\centering
\includegraphics[angle=90,width=0.4\textwidth]{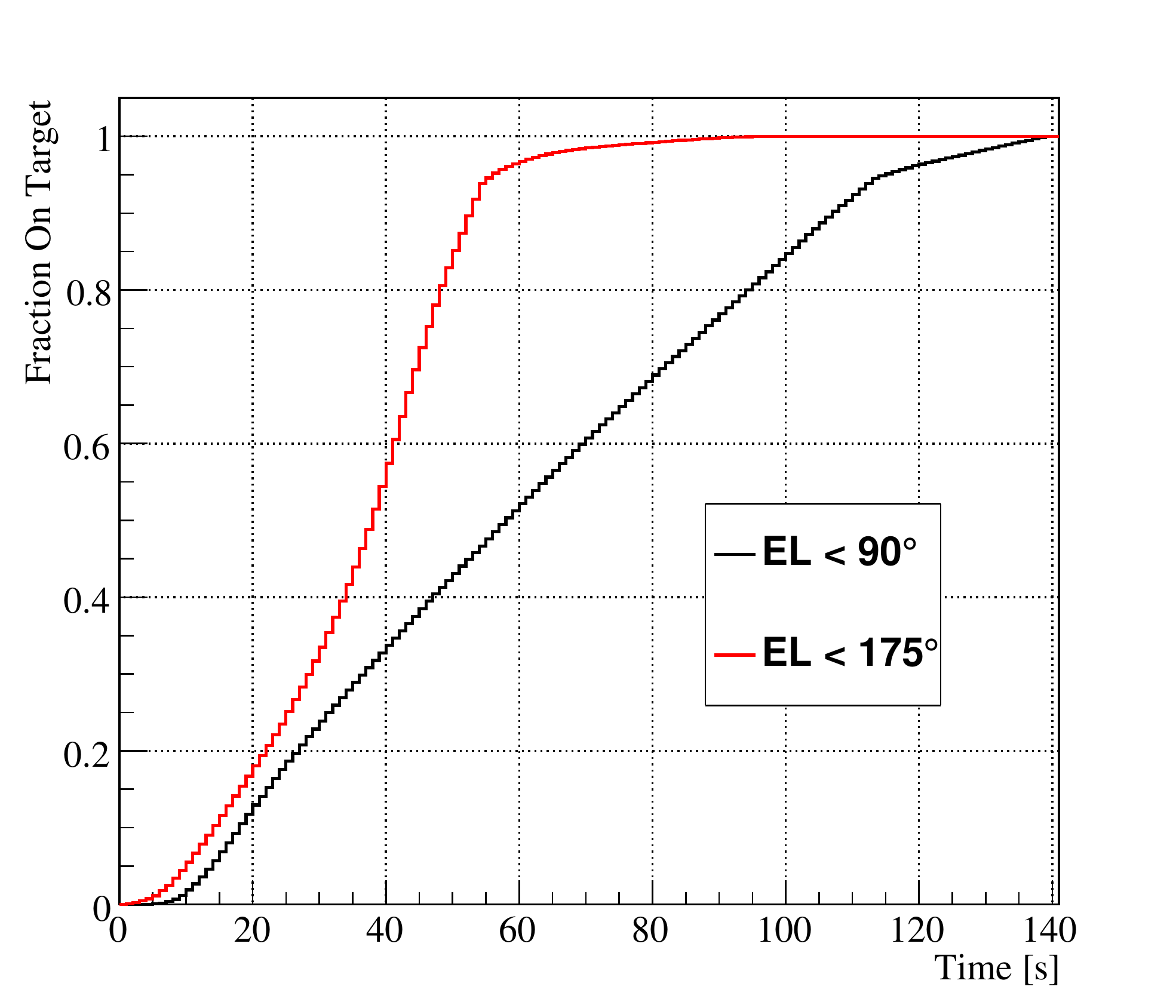}
\caption{The fraction of times the CT5 telescope is on target to a random position on the sky versus the time since the start of the repositioning. The fraction is shown with (EL$<$$175^{\circ}$, red line) and without (EL$<$$90^{\circ}$, black line) reverse observations enabled.}
\label{fig:timeontarget}
\end{figure}

A Monte-Carlo study was performed to estimate the time on target for the H.E.S.S. II telescope from a random observation position to a random position on the sky. In this study, the observation position was randomly sampled from the first $\sim$5 months of observations and the repositioning time for each axis for a given distance was taken from measurements using the real telescope. The actual motion range of the telescope was used, and the study was made both with and without reverse observations enabled. For the former case, the time on target was calculated to the point where the target is on the edge of the FoV of the telescope, as described above. The result of the study is shown in Fig.~\ref{fig:timeontarget}, and is summarized in Table~\ref{tab:timeontarget}.

\begin{table}[h]
\begin{center}
\begin{tabular}{l|cc}
Configuration & Mean time [s] & 90\% on target [s]\\ \hline
$\mathrm{EL}<90^{\circ}$   & 61  & 107 \\ 
$\mathrm{EL}<175^{\circ}$ & 36 & 52 \\ 
\end{tabular}
\caption{\emph{Column 2} The mean time on target for a repositioning to a random position on the sky. \emph{Column 3} The time needed for the telescope to be on target for 90\% of repositionings to random positions on the sky.}
\label{tab:timeontarget}
\end{center}
\end{table}
{\footnotesize
\section{Acknowledgments}
We appreciate the excellent work of the technical support staff in Heidelberg in the construction and operation of the equipment. We are also greatly in debt to the staff on site in Namibia, without who the commissioning and operation of the drive system would have never been possible. The assistance of MT Mechatronics during the commissioning is acknowledged.

}

\end{document}